\documentclass[12pt]{article}
\usepackage{subeqn}
\usepackage{epsfig}
\newcommand{\be}{\begin{equation}}
\newcommand{\ee}{\end{equation}}

\newcommand{\no}{\noindent}
\newcommand{\ce}{\begin{center}}
\newcommand{\nc}{\end{center}}

\makeatletter
\@addtoreset{equation}{section}
\makeatother

\def\sqr#1#2{{\vcenter{\vbox{\hrule height.#2pt
\hbox{\vrule width.#2pt height#1pt \kern#1pt
\vrule width.#2pt} \hrule height.#2pt}}}}

\def\operp{\hbox{${\kern+.25em{\bigcirc}
\kern-.85em\bot\kern+.85em\kern-.25em}$}}

\def\lsim{\;\raise0.3ex\hbox{$<$\kern-0.75em\raise-1.1ex\hbox{$\sim$}}\;}
\def\gsim{\;\raise0.3ex\hbox{$>$\kern-0.75em\raise-1.1ex\hbox{$\sim$}}\;}
\def\no{\noindent}

\def\ce{\centerline}
\def\ve{\vfill\eject}
\def\rdots{\mathinner{\mkern1mu\raise1pt\vbox{\kern7pt\hbox{.}}\mkern2mu
\raise4pt\hbox{.}\mkern2mu\raise7pt\hbox{.}\mkern1mu}}

\def\e e{$e^+ e^-$ }

\begin{document}

\ce{\bf A KNOT MODEL SUGGESTED BY THE}
\ce{\bf STANDARD ELECTROWEAK THEORY}
\vskip.5cm

\ce{\it Robert J. Finkelstein}
\vskip.3cm

\ce{Department of Physics and Astronomy}
\ce{University of California, Los Angeles, CA 90095-1547}
\vskip1.0cm

\no{\bf Abstract.}  We attempt to go beyond the standard electroweak theory
by replacing $SU(2)$ with its $q$-deformation:
$SU_q(2)$.  This step introduces new degrees of freedom that
we interpret as indicative of non-locality and as a possible basis for
a solitonic model of the elementary particles.  The solitons are conjectured
to be knotted flux tubes labelled by the irreducible representations of
$SU_q(2)$, an algebra which is not only closely related to the standard 
theory but also plays an underlying role in the description of knots.  
Each of the four
families of elementary fermions is conjectured to be represented by one
of the four possible trefoils.  The three individual fermions belonging
to any family are then assumed to occupy the three lowest states in the excitation spectrum
of the trefoil for that family.  One finds a not unreasonable variation of $q$
among the lepton and quark families.  The model in its present form predicts
a fourth generation of fermions as well as a neutrino mass spectrum.  The model
may be refined depending on whether or not the fourth generation is found.

\vskip.5cm

\no {\bf UCLA/04/TEP/30}

\vskip1.0cm

\no PACS \#s:
\vskip.3cm

\indent 02.20 Uu \\
\indent 02.10 Kn \\
\indent 12.60 Fr 

\ve

\section{Introduction.}

To go beyond the standard electroweak theory one may attempt to replace its
symmetry group: $SU(2)_L\times U(1)$.  In particular the local gauge
theory based on the quantum group $SU_q(2)_L$ is an attractive
possibility since its linearization agrees closely with the standard
theory in lowest order,$^1$ and since its $q=1$ limit is the standard
theory.  Moreover it has the required additional degrees of freedom if solitons
are to replace the point particles of standard theory in a more realistic
picture.

$SU_q(2)$ may be defined by the two-dimensional representation of any
member, $T$, as follows:
\begin{eqnarray}
T^t\epsilon_qT &=& T\epsilon_qT^t = \epsilon_q \\
T^\dagger &=& T^{-1} 
\end{eqnarray}
\be
\epsilon_q = \left(\matrix{0 & q_1^{1/2} \cr
-q^{1/2} & 0 \cr} \right) \qquad q_1 = q^{-1}
\ee
\no where $q$ is a real dimensionless number intended as an effective
measure of the new degrees of freedom necessary to describe the solitons.

If one sets
\be
T = \left(\matrix{a & b \cr -q_1\bar b & \bar a \cr} \right)
\ee
\no then by (1.1)-(1.3)
\be
\begin{array}{ll}
ab = qba & a\bar a + b\bar b = 1 \\
a\bar b = q\bar ba & \bar aa + q_1^2\bar bb = 1 \\
b\bar b = \bar bb & 
\end{array}
\ee
\no If $q\not = 1$ there are no finite matrix representations of this
algebra unless $q$ is a root of unity, which we exclude.

The $(2j+1)$ dimensional irreducible representation of $SU_q(2)(D^j_{mm^\prime}(a,\bar a,b,\bar b))$ resembles closely the $(2j+1)$ dimensional
irreducible representation of $SU(2)(D^j_{mm^\prime}(\alpha\beta\gamma))$ with
integers replaced by ``basic integers" and commuting arguments replaced
by $(a\bar ab\bar b)$.  The $(2j+1)$ dimensional irreducible representation
of $SU_q(2)$ is
\be
\begin{array}{rcl}
D^j_{mm^\prime}(a,\bar a,b,\bar b) &=& \Delta^j_{mm^\prime} \sum_{s,t}
\left\langle\matrix{n_+ \cr s}\right\rangle_1
\left\langle\matrix{n_- \cr t}\right\rangle_1
q^{t(n_++1-s)}(-1)^t\delta(s+t,n^\prime_+) \\
& &\times a^sb^{n_+-s}\bar b^t\bar a^{n_--t} \\
\end{array}
\ee
\no where
\be
\begin{array}{rcl}
n_\pm &=& j\pm m \\ n^\prime_\pm &=& j\pm m^\prime \\
\end{array} \quad
\left\langle\matrix{n \cr s}\right\rangle_1 =
{\langle n\rangle_1!\over \langle s\rangle_1!\langle n-s\rangle_1!}
\quad \langle n\rangle_1 = {q_1^{2n}-1\over q_1^2-1} \nonumber
\ee 
\be
\Delta^j_{mm^\prime} = \left[{\langle n^\prime_+\rangle_1!~
\langle n^\prime_-\rangle_1!\over
\langle n_+\rangle_1!~\langle n_-\rangle_1!}\right]^{1/2} \qquad
q_1 = q^{-1} \nonumber
\ee

\section{The Soliton Fields.}

We assume that all quantum fields lie in the (1.5) algebra and
on top of the usual expansions they are all 
expanded in irreducible representations of $SU_q(2)$ (a complete orthonormal 
set here designated as $D^j_{mm^\prime}(a,\bar a,b,\bar b))$.  Then the normal
modes, besides describing states of momentum and spin, will also
contain the factor $D^j_{mm^\prime}(a,\bar a,b,\bar b)$.  The
$D^j_{mm^\prime}(a,\bar a,b,\bar b)$ are polynomials in the non-commuting arguments
$(a,\bar a,b,\bar b)$ that obey the algebra (1.5) and have expectation
values that may be computed on the state space attached to the algebra.
Since $[b,\bar b]=0$, the eigenstates of $b$ and $\bar b$ may be chosen
as basic states.  Then $a$ and $\bar a$ are lowering and raising operators.
Set
\be
\begin{array}{rcl}
b|0\rangle &=& \beta|0\rangle  \nonumber \\
\bar b|0\rangle &=& \beta^\star|0\rangle \nonumber
\end{array}
\ee
\no Then
\be
\bar b\bar a^n|0\rangle = q^n\beta^\star\bar a^n|0\rangle
\ee
\no or
\be
\begin{array}{rcl}
\bar b|n\rangle &=& q^n\beta^\star|n\rangle  \nonumber\\
b\bar b|n\rangle &=& q^{2n}|\beta|^2|n\rangle \nonumber
\end{array}
\ee
\no The expectation values $\langle n|D^j_{mm^\prime}(a,\bar a,b,\bar b)|n\rangle$
are polynomials in $q$ and $|\beta|^2$.  The states of excitation $|n\rangle$
of the objects represented by $D^j_{mm^\prime}$ are analogous to the excited states
of a string.

\ve

Extending current ideas we may ask whether the different normal modes
$(jmn)$ represent strings, loops, and knots in 3 dimensions.  Because
they are representative of $SU_q(2)$ rather than $SU(2)$, all
modes have an internal excitation spectrum and in this sense may be interpreted as solitons.

The $SU_q(2)$ symmetry group suggests additional features of these
solitons since the Kauffman algorithm$^2$ for associating a Jones polynomial
with a knot may be expressed in terms of $\epsilon_q$ 
alone,$^3$ where $\epsilon_q$ is also the basic invariant of $SU_q(2)$
as shown in (1.1).  We are thus led to label physical knots by the 
irreducible representations of $SU_q(2)$ that we now express as
$D^{N/2}_{{w\over 2}{r+1\over 2}}$ where $(N,w,r)$ mean the number of
crossings, the writhe, and the rotation of the knot.  This step is 
certainly permissible and is clearly
suggested by the $SU_q(2)$ symmetry.  Moreover gauge fields are known to
exhibit flux tubes and these may be knotted, as is also shown in the classical
limit by the work of Fadeev and Niemi.$^4$  One knows as well that there
may be knots of magnetic flux in the classical Maxwell field.  One may then
conjecture that some but not all of the solitonic normal modes are knotted flux
tubes labelled by $D^{N/2}_{\frac{w}{2}\frac{r+1}{2}}$.

\section{Assignment of Particle States to $D^j_{mm^\prime}(a,\bar a,b,\bar b)$.}

We would like to associate the knotted solitons with the
observed point particles and therefore we shall describe the simplest knots
by the same quantum numbers that characterize the leptons and quarks, namely, $t,t_3$ and either the hypercharge $t_0$ or the charge
$Q$ related by $Q = t_3+t_0$.  We wish to connect $t,t_3$ and $Q$ with
the three knot labels $(N,w,r)$.

We shall represent the elementary Fermions (leptons and quarks) by the
simplest knots (trefoils).  The following table describes a possible
relation between $(t,t_3,Q)$ and $(N,w,r)$.

In Table 1 we have listed possible knot assignments for leptons, quarks,
and Higgs as well as the relation between their knot labels and their
labels in the standard theory.

\ve

\vskip 0.1in
\begin{center}{{\it Left-Handed Lepton States}} 
\end{center}
\no
\begin{tabular}{l|cccc|ccc|c}
   & $t$ & $t_{3}$ &  $t_{0}$& Q & $N$ & $w$ & $r$ & 
$D^{N/2}_{w/2  \;\;(r+1)/2} $\\
\hline
$ 
l_{L} \left( \begin{array}{c}
               \nu_{L} \\
                e_{L} \\
             \end{array} \right)   
\left( \begin{array}{c}
               \nu^{'}_{L} \\
                \mu_{L} \\
             \end{array} \right)   
\left( \begin{array}{c}
               \nu^{''}_{L} \\
                \tau_{L} \\
             \end{array} \right) $ &
$\begin{array}{c}
           \frac{1}{2}\\
           \frac{1}{2}\\
    \end{array} $ &         
$\begin{array}{c}
           \frac{1}{2}\\
           -\frac{1}{2}\\
    \end{array} $ &         
$\begin{array}{c}
           -\frac{1}{2}\\
           -\frac{1}{2}\\
    \end{array} $ &         
$\begin{array}{c}
           0\\
           -1\\
    \end{array} $ &         
$\begin{array}{c}
           3\\
           3\\
    \end{array} $ &         
$\begin{array}{c}
           3\\
           -3\\
    \end{array} $ &         
$\begin{array}{c}
           2\\
           -2\\
    \end{array} $ &         
$\begin{array}{c}
          D^{3/2}_{3/2 \;\; 3/2} \\
          D^{3/2}_{-3/2 \;\;  -1/2} \\
    \end{array} $          
\end{tabular}

\vskip 0.2in
\begin{center}{{\it Left-Handed Quark States}} 
\end{center}
\no
\begin{tabular}{l|cccc|ccc|c}
   & $t$ & $t_{3}$ &  $t_{0}$& Q & $N$ & $w$ & $r$ & 
$D^{N/2}_{w/2  \;\;(r+1)/2} $\\
\hline
$ 
q_{L} \left( \begin{array}{c}
               u_{L} \\
                d_{L} \\
             \end{array} \right)   
\left( \begin{array}{c}
               c_{L} \\
                s_{L} \\
             \end{array} \right)   
\left( \begin{array}{c}
               t_{L} \\
               b_{L} \\
             \end{array} \right) $ &
$\begin{array}{c}
           \frac{1}{2}\\
           \frac{1}{2}\\
    \end{array} $ &         
$\begin{array}{c}
           \frac{1}{2}\\
           -\frac{1}{2}\\
    \end{array} $ &         
$\begin{array}{c}
           \frac{1}{6}\\
           \frac{1}{6}\\
    \end{array} $ &         
$\begin{array}{c}
           \frac{2}{3}\\
           -\frac{1}{3}\\
    \end{array} $ &         
$\begin{array}{c}
           3\\
           3\\
    \end{array} $ &         
$\begin{array}{c}
           3\\
           -3\\
    \end{array} $ &         
$\begin{array}{c}
           2\\
           -2\\
    \end{array} $ &         
$\begin{array}{c}
          D^{3/2}_{3/2 \;\; 3/2} \\
          D^{3/2}_{-3/2 \;\;  -1/2} \\
    \end{array} $          
\end{tabular}

\vskip 0.2in
\begin{center}{{\it Higgs}}
\end{center}
\begin{center}{
\begin{tabular}{l|cccc|ccc|c}
   & $t$ & $t_{3}$ &  $t_{0}$& Q & $N$ & $w$ & $r$ & 
$D^{N/2}_{w/2  \;\;(r+1)/2} $\\
\hline
$ 
\mbox{Higgs} \;\;   \left( \begin{array}{c}
               \phi_{+} \\
                \phi_{0} \\
    \end{array} \right) $ &         
$\begin{array}{c}
           \frac{1}{2}\\
           \frac{1}{2}\\
    \end{array} $ &         
$\begin{array}{c}
           \frac{1}{2}\\
           -\frac{1}{2}\\
    \end{array} $ &         
$\begin{array}{c}
           \frac{1}{2}\\
          \frac{1}{2} \\
    \end{array} $ &         
$\begin{array}{c}
           1\\
           0 \\
    \end{array} $ &         
$\begin{array}{c}
           3\\
           3\\
    \end{array} $ &         
$\begin{array}{c}
           3\\
           -3\\
    \end{array} $ &         
$\begin{array}{c}
           2\\
           -2\\
    \end{array} $ &         
$\begin{array}{c}
          D^{3/2}_{3/2 \;\; 3/2} \\
          D^{3/2}_{-3/2 \;\;  -1/2} \\
    \end{array} $     
\end{tabular}
}
\end{center}
    
\vskip 0.2in
\begin{center}{ 
{\it Relations between Conventional Labels and Knot Labels} 
}
\end{center}

\begin{center}{
\begin{tabular}{|l|l|l|}
$l_{L}$ & $q_{L} $ & Higgs \\
\hline
$ \begin{array}{l}
            t=\frac{N}{6} \\
            t_{3}=\frac{w}{6} \\
            Q=\frac{1}{4}r-\frac{1}{2} \\
       \end{array} $ &
$  \begin{array}{l}
            t=\frac{N}{6} \\
            t_{3}=\frac{w}{6} \\
            Q=\frac{1}{4}r+\frac{1}{6}\\
       \end{array} $ &
$ \begin{array}{l}
            t=\frac{N}{6} \\
            t_{3}=\frac{w}{6} \\
            Q=\frac{1}{4}r+\frac{1}{2} \\
       \end{array} $ 
\end{tabular}
}
\end{center}
\no
Here $t_{0}=Q-t_{3}$. The left handed components are doublets
and the right handed components are singlets. 
The chiral trefoils are associated with the higher charge states.

\begin{center}{\bf Table 1.}
\end{center}

\ve

\no This table represents only an illustrative choice.  The real question,
which is discussed in the next paragraph, is whether the empirical information favors a particular choice.

\section{Mass Spectra.}

Motivating the present work is the conjecture that some essential
features of the elementary fermionic
solitons can be represented in the solitonic picture suggested
by $SU_q(2)\times SU(1)$, since the linearization of the $q$-theory
agrees very well with the standard electroweak theory and its $q=1$
limit is standard theory.

There are two obvious ways in which this picture can be tested.
The first is to study the systematics of reaction rates computed on the
one hand between the solitons and on the other between the corresponding
point particles.  We discuss this program elsewhere.$^5$

The second is to ask whether the spectra of the solitons in this model
are reasonable.  Here there is a preliminary result obtained by 
simultaneously testing the old idea that the muon is an excited state
of the electron.  We shall extend this idea by first arranging the
elementary Fermions in the following four families.
\be
\begin{array}{llll}
{\rm (1)} & e & \mu & \tau \cr
{\rm (2)} & d & s & b \cr
{\rm (3)} & u & c & t \cr
{\rm (4)} & \nu_e & \nu_\mu & \nu_\tau 
\end{array}
\ee
\no Members of the same family share the same quantum numbers
$(t,t_3,Q)$ as shown in the table.  We now ask whether the three
members of each family can be identified with the ground and first two
excited states of a soliton.  Assuming that this is possible, let us
then assume that each family is represented by a single soliton.  Then
there are 4 solitons, one for each family.  There are also only 4 trefoils
and these may be matched against the 4 solitons.  To test a given match
one needs to calculate the spectrum of excited states of each soliton
(trefoil) and compare with the empirical mass spectrum of each family.

To calculate the mass spectrum of a soliton, we follow the standard
theory to the extent of assuming a mass term of the following form
\be
{\cal{M}} \sim (\bar\psi_L\varphi_0\psi_R + \bar\psi_R\varphi_0\psi_L) 
\ee
\no where $\varphi_0$ is the neutral component of the Higgs doublet in
the unitary gauge and $\psi$ is a lepton or quark field.  We have been
assuming that all fields including the Higgs field, and therefore the Higgs
potential, lies in the $q$-algebra (1.5).  Now let the Higgs potential be
so chosen that its minima be at the trefoil points.

Now replace
$\psi_L$ and $\varphi_0$ by the normal modes of these fields that
represent trefoils.  Since $\psi_R$ is a singlet in the standard
theory, we assume that it is also a singlet in the $SU_q(2)$ theory.
Then within the $SU_q(2)$ algebra
\be
{\cal{M}} \sim \bar\psi_L\varphi_0 + \psi_L\varphi_0
\ee
\no where $\psi_L$ and $\varphi_0$ are chosen from any of the 4 trefoil
representations of $SU_q(2)$.

Here are the four trefoils with the following writhe and rotation
$(w,r)$
\be
(3,2) \qquad (3,-2) \qquad (-3,-2) \qquad (-3,2)
\ee
\no The knot labels $(D^{N/2}_{{w\over 2}{r+1\over 2}})$ with their
dependence on the $SU_q(2)$ algebra according to Eq. (1.6) are listed as
follows:
\be
\begin{array}{cccc}
\mbox{I} & \mbox{II} & \mbox{III} & \mbox{IV}\\
D^{3/2}_{3/2 \; 3/2} \sim a^{3} &
D^{3/2}_{3/2 \; -1/2} \sim a b^{2} &
D^{3/2}_{-3/2 \; -1/2} \sim \bar{a}^{2} \bar{b} &
D^{3/2}_{-3/2 \; 3/2} \sim \bar{b}^{3} \\
(3,2) & (3,-2) & (-3,-2) & (-3,2) \\
\end{array}
\ee
\no Then the mass operator (4.3) associated with any soliton $(w,r)$
and Higgs $(w^\prime r^\prime)$ becomes
\be
{\cal{M}}(w,r;w^\prime,r^\prime) \sim \bar D^{3/2}_{{w\over 2}{r+1\over 2}}
D^{3/2}_{{w^\prime\over 2}{r^\prime+1\over 2}}
+ D^{3/2}_{{w\over 2}{r+1\over 2}}D^{3/2}_{{w^\prime\over 2}{r^\prime+1\over 2}}
\ee
\no The expectation value of ${\cal{M}}(w,r;w^\prime r^\prime)$ vanishes 
unless $w=w^\prime$ and $r=r^\prime$.  Then
\be
\langle n|{\cal{M}}(w,r)|n\rangle \sim
\langle n|\bar D^{3/2}_{{w\over 2}{r+1\over 2}}
D^{3/2}_{{w\over 2}{r+1\over 2}}|n\rangle
\ee

To accommodate the 4 families one needs 4 minima in the Higgs potential.
These minima may be labelled by the magnitudes of the Higgs field $\varphi_0$
and by the associated Higgs trefoils.  The mass scale of each family is
determined by $\varphi_0$ at the minimum for that family, and the trefoil for
that family must agree with the trefoil for $\varphi_0$.  With this
understanding Eq. (4.7) implies
\be
m_n(w,r) \sim \langle n|\bar D^{3/2}_{{w\over 2}{r+1\over 2}}
D^{3/2}_{{w\over 2}{r+1\over 2}}|n\rangle
\ee
\no where $m_n(w,r)$ is the mass of the $(w,r)$ soliton at the n$^{\rm th}$
level.  The different spectra corresponding to the different solitons
(I, II, III, IV) in (4.5) are given by
\be
\begin{array}{lcl}
~~\mbox{I} \;\;\;\bar{a}^{3} a^{3}|n> & = & 
[(1-q^{2n-2} |\beta|^{2})(1-q^{2n-4} |\beta|^{2})(1-q^{2n-6} |\beta|^{2})]|n\rangle \\ 
~\mbox{II} \;\;\; \bar{b}^{2} \bar{a} a b^{2}|n> & = & 
[q^{4n} |\beta|^{4}-q^{6n-2} |\beta|^{6} ]|n\rangle\\ 
\mbox{III} \;\;\; b a^{2} \bar{a}^{2} \bar{b}|n> & = & 
[(q^{2n} |\beta|^{2})(1-q^{2n} |\beta|^{2})(1-q^{2n+2} |\beta|^{2})]|n\rangle \\
~\mbox{IV} \;\;\; b^{3} \bar{b}^{3}|n> & = & 
[q^{6n} |\beta|^{6}]|n\rangle \\ 
\end{array}
\ee

We shall now regard the 3 particles of one family as ground and 
first two excited states of the same soliton,
Denote the ratios of these three masses by
 
\be
\begin{array}{lcl}
M &=& \frac{<1|{\cal M}|1>}{<0|{\cal M}|0>} \\
m &=& \frac{<2|{\cal M}|2>}{<1|{\cal M}|1>} \\
\end{array}
\ee
\no Note that we have not assumed $a|0\rangle =0$.

In the above 4 cases we find:

\be
\begin{array}{lcl}
\mbox{I} \qquad \qquad \frac{m-1}{m-q^{6}} &=& q^{2} \frac{M-1}{M-q^{6}} \\
\end{array}
\ee

\be
\begin{array}{lcl}
\mbox{II} \qquad \qquad \frac{m-q^{4}}{m-q^{6}} &=& q^{2} \frac{M-q^{4}}{M-q^{6}} \\
\end{array}
\ee

\be
\begin{array}{lcl}
\mbox{III} \qquad \qquad \frac{m-q^{2}}{m-q^{6}} &=& q^{2} \frac{M-q^{2}}{M-q^{6}} \\
\end{array}
\ee

\be
\begin{array}{lcl}
\mbox{IV} \qquad \qquad M=m &=& q^{6} \\
\end{array}
\ee

For the 4 families we have

\be
\begin{array}{lcc}
 &  M & m \\
(1) \qquad \qquad e,\mu,\tau & 193 & 16.7\\
(2) \qquad \qquad d,s,b & 37.5 & 31.8 \\
(3) \qquad \qquad u,c,t & 750 & 117 \\
(4) \qquad \qquad \nu_{e}, \nu_{\mu}, \nu_{\tau} & ? & ? \\
\end{array}
\ee

These ratios depend on the following estimates of the masses of
the quarks
\be
\begin{array}{ccccccc}
u & d & c & s & t & b & \\
.002 & .004 & 1.5 & .15 & 176 & 4.7 & \mbox{GeV/c}^2
\end{array}
\ee

One notes that the (1), (2) and (3) families can not be assigned to the IV
spectrum since their masses are not in geometric progression, i.e., 
$M \not= m$.
Then the neutrino family would have to be assigned to IV, and the masses
of the three neutrinos would then be expected to be in geometric progression.

To decide how to match the families ((1), (2), (3)) with the spectra
(I, II, III), the Eqs. (4.11)-(4.13) can be rewritten as
\be
\begin{array}{lcl}
\mbox{I} \;\;\; x^{4} - r x^{3} -m x + r M &=& 0\\
\end{array}
\ee
\be
\begin{array}{lcl}
\mbox{II} \;\;\; x^{6} - x^{5}-M x^{4} + M x^{2}
+m M x -m M &=& 0\\
\end{array}
\ee
\be
\begin{array}{lcl}
\mbox{III} \;\;\; x^{5}-(M+1)x^{4}+mx^{3}-mx^{2}+M(m+1)
x-mM &=& 0\\
\end{array}
\ee
 
\no
Here $x=q^{2}$, $r=(m-1)/(M-1)$.

Since $x=q^{2}$, we are interested only in positive roots. Equation
(I) has at most 2 positive roots, (II) has at most 3 and (III) has at most 5.
In every case $q=1$ is a root but is uninteresting. The number of interesting positive
roots is then (1,2, and 4) in these three cases.
 
Of the three families $(e,\mu,\tau)$ has the least structure since these particles do not
have hypercharge or gluon charge. For this reason we assign  $(e,\mu,\tau)$
to I. The solution
of I with $M=193$ and $m=16.7$ is
\be
(e,\mu,\tau) \;\;\;\;\; q=1.46 \;\;\;\;\; |\beta| = 3.20
\ee
  
The other two cases $(d,s,b)$ and $(u,c,t)$ are rather arbitrarily assigned to (II)
and (III) respectively and in each case we choose the minimum root for $q$.
One finds

\be
(d,s,b) \;\;\;\;\;  q=1.76 \;\;\;\;\;  |\beta| = 3.33
\ee
\be
(u,c,t) \;\;\;\;\;  q=2.12 \;\;\;\;\;  |\beta| = 1.06
\ee

The neutrino family is compatible with any value of $q$.

The values of $q$ in (4.19)-(4.22) are close to the unperturbed value 
($q=1$) for the standard theory
and do not vary much between the minima in the Higgs potential.  The main uncertainty in these numerical results comes
from the meaning of the estimated ``masses" of $u$ and $d$.
Alternatively one could define the $u$ and $d$ masses by this model.

We regard the deformation parameter $q$ as an effective measure of 
external influences on the standard model.  Consistent with this view,
$q$ is not far from unity and the lepton family, having no gluon charge,
has a $q$ value closer to unity than the quark families.

We have assumed that the three observed particles of each family occupy the
3 lowest states of the soliton representing that family.  The model also
permits higher excited states but if these lie at very high energies, they
may have such short lifetimes that they would not be observable as particles.
The tentative assignment that we have assumed in Table 1 leads to a fourth
generation of (-1/3 quarks) at $30 m_b \sim 144$ GeV and a fourth generation
of (2/3 quarks) at $94.6 m_t \sim 16,650$ GeV.  The corresponding fourth
generation lepton would appear at the $12 m_\tau \sim 21.3$ GeV which is excluded by the known decays of the $Z^0$.  If the
assignments of $dsb$ and $uct$ are interchanged in (4.9) so that $dsb$
corresponds to III and $uct$ to II then the fourth generation would appear
at $30.1 m_b$ and $106 m_t$.  If a fourth generation should be observed
then a unique assignment of the $(dsb)$ and $(uct)$ families to trefoils could
be put on an empirical basis.  Depending on whether or not a fourth generation
is observed, the model may be improved.

\vskip.5cm

\no {\bf Remarks.}

One may in principle construct a field theory based on normal mode
expansions in the irreducible representations of $SU_q(2)$ [instead of
an expansion in the generators of the Lie algebra of $SU(2)]$.  
The linearization of this
theory would approach the standard theory in its $q=1$ limit and is
the motivation for the present note.  We have not developed this field
theory here but have discussed some of the qualitative features that might
be expected of it.  The resulting solitonic model is sufficiently close
to the point particle model of the standard theory to be of possible
interest as a phenomenological model for organizing facts not accessible
from the standard theory.  The model as here presented depends
on a number of simplifying assumptions.  If these are accepted the model
would also predict the mass ratios of the neutrinos and a fourth generation of
fermions.  Depending on whether or not these predictions are approximately
confirmed, the simplifying assumptions may be dropped and the model may be
refined.

\vskip 0.1in

\no
{\bf References.} 

\vskip 0.1in
1. R. J. Finkelstein, Lett. Math. Phys. {\bf 62}, 199 (2002).

2. L. H. Kauffman, Int. J. Mod. Phys. A{\bf 5}, 93 (1990).

3. R. J. Finkelstein, hep-th/0311192.

4. L. Fadeev and A. J. Niemi, hep-th/9610193.

5. A. C. Cadavid and R. J. Finkelstein, to appear.

\end{document}